\documentclass[10pt,conference]{IEEEtran}
\usepackage{epsfig,setspace,amsmath,epsf,amssymb,bm,theorem,cite,graphicx, epstopdf,algorithm,float,color,mathtools, authblk,physics}
\usepackage[table,xcdraw]{xcolor}
\usepackage{subcaption}
\usepackage{algorithm,algorithmic}
\usepackage{dsfont}

\DeclareMathOperator*{\argmax}{arg\,max}

\newtheorem{theorem}{Theorem}
\newtheorem{corollary}{Corollary}

\newtheorem{remark}{Remark}
\newtheorem{lemma}{Lemma}

\newcommand{\e}{{\mathbb{E}}}

\newcommand{\custmsz}{0.89}

\IEEEoverridecommandlockouts
\allowdisplaybreaks

\begin{document}




\title{Utilizing the Perceived Age to Maximize Freshness in Query-Based Update Systems}

\author[1]{Sahan Liyanaarachchi}
\author[1]{Sennur Ulukus}
\author[2]{Nail Akar}

\affil[1]{\normalsize University of Maryland, College Park, MD, USA}
\affil[2]{\normalsize Bilkent University, Ankara, T\"{u}rkiye}

\maketitle
\begin{abstract}
  Query-based sampling has become an increasingly popular technique for monitoring Markov sources in pull-based update systems. However, most of the contemporary literature on this assumes an exponential distribution for query delay and often relies on the assumption that the feedback or replies to the queries are instantaneous. In this work, we relax both of these assumptions and find optimal sampling policies for monitoring continuous-time Markov chains (CTMC) under generic delay distributions. In particular, we show that one can obtain significant gains in terms of mean binary freshness (MBF) by employing a waiting based strategy for query-based sampling.
\end{abstract}

\section{Introduction}
With the establishment of AI factories in the new age of computing supremacy, there is an increasing need to properly monitor the available computing resources. For this purpose, metrics such as age of information (AoI) \cite{yates2020age, age1, age2, rts2012, rtsms2012}, age of incorrect information (AoII) \cite{AoII2019, AoII_Markov} and binary freshness (BF) \cite{melih_BF_cache, melih_BF_gossip, melih_BF_Inf, melih_IF_CUS, graves2024} have been well-established throughout the literature. Among them, one of the most widely used metrics when monitoring Markov sources is the BF metric as it is a direct proxy for the probability of error of the estimates. If $X(t)$ is the Markov source and $\hat{X}(t)$ is its estimate at the monitor, then the mean binary freshness (MBF), denoted by $\e[\Delta]$, is defined as,
\begin{align}
    \e[\Delta]=\limsup_{T\to\infty}\frac{1}{T}\e\left[\int_0^T\mathds{1}\{X(t)=\hat{X}(t)\}\dd{t}\right].
\end{align}

When monitoring Markov sources in a remote estimation setting, a common practice is to query the state of the source rather than for the source to constantly send its update to the remote monitor. For example, consider a GPU cluster shared by many users where the users would need to know the status of the GPU cluster, whether it is free or occupied, before submitting jobs to it. In such scenarios, where there is a large population of users connected to one cluster, it is common for the users to inquire the status of the cluster rather than for the cluster to broadcast its status to every user. Query-based sampling is a natural fit for such a setting.

\begin{figure}
    \centering
    \includegraphics[width=0.9\columnwidth]{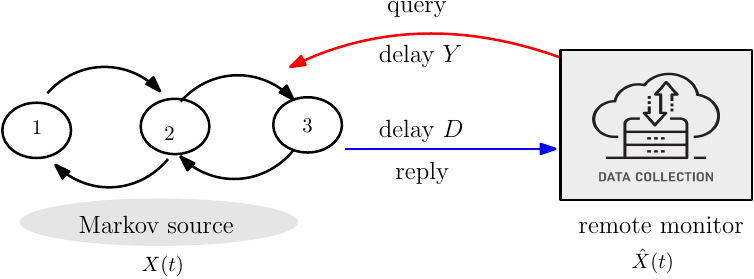}
    \caption{System model.}
    \label{fig:sys_model}
\end{figure}

In query-based sampling, the remote monitor sends queries to the source and upon receiving a query, the source replies back with its status (see Fig.\ref{fig:sys_model}). Most of the prior works in this domain are confined to exponentially distributed inter-query times and instantaneous replies (i.e., feedback). Additionally, in most cases, a martingale estimator, which uses the most recently received update as its current estimate, is used. In this work, we break away from these restrictions and relax these assumptions to find the optimal sampling strategy to maximize the MBF metric when employing a query-based sampling to monitor a CTMC. In particular, we realize these sampling strategies for generic delay distributions, where we consider that both the query and its subsequent reply are subjected to arbitrary and random delays.

In this context, waiting-based sampling policies have been widely pursued under AoI and related metrics. However, its study has been more or less limited to push-based systems where the source is in control and sends its updates in its own accordance. For such systems, where the source has full information about the system, waiting based policies have been shown to be optimal. In this work, we show that, the same strategy is optimal even when we only have an outdated view of the system. Through rigorous proofs and simulations, we show that one can obtain significant gains in terms of MBF metric by waiting for some time before sending out queries upon receiving a reply from the CTMC. 

\section{Related Work}
In \cite{nail_QS}, query-based sampling procedure was used to monitor multiple CTMCs under a martingale estimator. In there, the authors assumed that querying times are exponentially distributed so as the preserve the Markov structure of the problem and also assumed that replies were instantaneous. Under these assumptions, they find the optimal rate allocation policy to maximize several variants of the BF. This work was later extended to non-martingale estimators in \cite{structured_estimators}. The works in \cite{Markov_machines} and \cite{revMax} adapt query-based sampling under similar assumptions to maximize the utility of Markov machines.

For evaluating pull-based update systems, two new metrics known as effective age of information (EAoI) and query age of information (QAoI) were introduced in \cite{EAoI} and \cite{QAoI}, respectively. These two metrics take samples out the AoI process at query times and then aim to minimize the average of the sampled AoI process. This was later extended in \cite{QVAoI} by incorporating semantic aspects such as version age into the QAoI metric. In all these works, the sources decide when to transmit data based on the querying process, and hence, lack the true essence of a pull-based system.

Waiting policies have been shown to be optimal for minimizing AoI and AoII metrics when transmitting status updates over a random delay channel \cite{ys17,ys19,AoIIDelay,lts2015}. Most of these works rely on an instantaneous feedback channel. Only a very few works in this domain consider delays in the feedback channel as well \cite{twd2020, urtwd2022, early_sample}. However, in all these works, the control is at the source-side, where the source has the full information of the monitored process. The work in \cite{ismail_map} is one of the few avenues which looks into a pull-based system where the remote monitor is in control. They aim to minimize the AoII metric using a belief MDP, but their work assumes that query delays are negligible and reply delays are one time unit.

\section{System model}
Let $X(t)$ denote an irreducible finite-state CTMC with states $\mathcal{S}=\{1,2,\dots,S\}$ and let $\hat{X}(t)$ be its estimate maintained at the remote monitor. Let $Q$ be the generator matrix of the CTMC and $P(t)=e^{Qt}$ be its transition probability matrix whose elements we denote by $P_{i,j}(t)$ which corresponds to the probability $\mathds{P}(X(t)=j|X(0)=i)$. The state of this CTMC is tracked by the monitor with the use of queries where the CTMC, upon the arrival of a query, responds back to the remote monitor with its status. We consider a generic delay channel for the transmission and the reception of the queries and their respective replies. Thus, when a reply to a query is received by the remote monitor, it will be aged by an amount equal to the delay experienced in the reply channel. Once a reply is received by the remote monitor, the monitor will wait for some time before sending out the next query to the CTMC. We assume that this waiting time is a bounded function of two parameters: the most recently received reply/update and its age (backward delay). In particular, if the most recent reply indicated that $X(t)$ was in state $i$, $d$ time units ago, where $d$ is the delay experienced by the reply, the remote monitor will wait $W(i,d)$ time units before sending out the next query. If the wait time is independent of the observed state, we will represent it as $W(d)$ and if it is independent of the age, we will represent it as $W_i$ to simplify the notation.

Let $Y$ denote the random variable associated with transmission delay incurred when sending a query to the CTMC (forward delay) and let $D$ denote the random variable associated with delays experienced by the replies from the CTMC to the remote monitor (backward delay). We will represent the distributions of these random variables with $F^Y(y)$ and $F^D(d)$. Moreover, let $Z=Y+D$ be the combined delay whose distribution function is represented by $F^Z(z)$. Let $\mathcal{Y}$, $\mathcal{D}$ and $\mathcal{Z}$ denote the support of the above three random variables, respectively. We assume that forward delay $Y$ and backward delay $D$ are non-negative random variables which are independent from each other and have finite expectations. Based on the distribution of these random delays, we will be finding the optimal wait times that should be followed by the remote monitor. In particular, we will be finding the optimal stationary waiting policy (waiting function only depends on the observed state and its age) that maximizes the MBF.

\section{Mean Binary Freshness}
In this section, we present the main results for the MBF under generic delay distributions and estimators. 

\begin{theorem}\label{thrm:fresh_gen}
    The MBF under a stationary  waiting policy is ,
    \begin{align}
        \e[\Delta]=\frac{\e\left[\int\limits_{D'}^{D'+W(X,D')+Z}P_{X,\hat{X}(t)}(t) \dd{t} \right]}{\e[W(X,D')]+\e[Z]}, \label{eqn:stat_fresh}
    \end{align}
    where $X$ is a discrete random variable over the states $\mathcal{S}$ with probability mass function $\bm\phi=\{\phi_1,\phi_2,\dots,\phi_S\}$ and $D'$ is an independent copy of $D$. Further, $\phi_i$ is the unique solution that satisfies the following  equations,
    \begin{align}
        \sum_{i=1}^S\phi_i\tilde{P}_{i,j}&=\phi_j,\quad
        \sum_{i=1}^S\phi_i=1,
    \end{align}
    where $\tilde{P}_{i,j}=\int_D\int_YP_{i,j}(d+W(i,d)+y)\,\dd{F^Y(y)}\,\dd{F^D(d)}$.
\end{theorem}

\begin{corollary}\label{cor:fresh_gen}
      The MBF under a stationary waiting policy is,
    \begin{align}
        \e[\Delta]=\frac{\sum_{i=1}^S\phi_i\e[g^W(i)]}{\e[Z]+\sum_{i=1}^S\phi_i\e[W(i,D)]},\label{eqn:fresh_phi}
    \end{align}
    where $\e[g^W(i)]=\int_\mathcal{D}\int_{0}^{\infty}P_{i,\hat{X}(t+d)}(t+d)(1-F^{Z}(t-W(i,d)))\,\dd{t}\,\dd{F^D(d)}$ is the average portion of time the estimator was fresh between two successive query replies,  given that the most recent reply was from state $i$.
\end{corollary}

The proofs of Theorem \ref{thrm:fresh_gen} and Corollary \ref{cor:fresh_gen} are presented in Appendix \ref{appen:fresh_gen} and Appendix \ref{appen:cor_fresh_gen}, respectively. Theorem \ref{thrm:fresh_gen} is a consequence of the ergodicity of the system and Corollary \ref{cor:fresh_gen} follows immediately from Theorem \ref{thrm:fresh_gen} by conditioning on the random variable $X$ which models the sampled state. Next, we show in Lemma \ref{lem:state_ind} that, if the waiting policy is independent of the observed state, then $X$ has the same distribution as the stationary distribution $\bm \pi=\{\pi_1,\pi_2,\dots,\pi_S\}$ of $X(t)$. This result will be crucial for constructing our waiting policies analytically. The proof of Lemma \ref{lem:state_ind} is given in Appendix \ref{appen:lem:state_ind}.

\begin{lemma}\label{lem:state_ind}
    If the waiting time function $W(i,d)$ is independent of state $i$, then $\phi_i=\pi_i$.
\end{lemma}
Finally, by invoking Lemma \ref{lem:state_ind}, we give an expression for MBF under a zero-wait policy (i.e., $W(i,d)=0$) in Corollary \ref{cor:zero_wait}.
\begin{corollary}\label{cor:zero_wait}
    The average freshness under a zero-wait policy for a generic delay distribution is given by,
    \begin{align}
        \e[\Delta]=\frac{\sum_{i=1}\pi_i\e[g(i)]}{\e[Z]},
    \end{align}
     where $\e[g(i)]=\e[g^W(i)]$ with $W(i,d)=0$.
\end{corollary}
\section{ Waiting Policies}\label{sec:waiting_policies}
In this section, we propose a semi-Markov decision process (SMDP) formulation to find optimal waiting function $W(i,d)$. However, finding the optimal waiting period across all possible states and ages based on the most recent update is an arduous task due to the curse of dimensionality that is inherent to SMDP formulations. Therefore, we will first look at two simplified waiting policies that reduce the state space along with a greedy waiting policy, and postpone the discussion of optimal waiting policy to the latter half of this section. To simplify the state space, we will be considering two perspectives, one where the waiting policy is independent of the observed state and the other where the waiting policy is independent of the perceived age of the received reply. 

\subsection{State-Independent Waiting Policy}\label{sec:state_ind}
In here, we assume that the waiting function is independent of the received state $X$ and only depends on the backward delay $D$. Despite the large state space  which is possibly continuous based on the distribution of $D$, this assumption enable us to invoke Lemma \ref{lem:state_ind} and solve for the waiting times analytically. To proceed with the analysis, we will first linearize the fractional objective function in \eqref{eqn:stat_fresh} using the Dinkelbach method \cite{dinkelbach}, as follows,
\begin{align}
    J(\theta)=\e\left[\int\limits_{D'}^{D'+W(D')+Z}\hspace{-5mm}P_{X,\hat{X}(t)}(t) \dd{t} \right]-\theta\e[W(D')+Z],
\end{align}
Let $\theta^*$ be the value of fractional objective function under the optimal state independent waiting policy and let $J^*(\theta)=\max_{W(\cdot)}{J(\theta)}$. It is well-known that $J^*(\theta)\lesseqqgtr 0 \iff \theta^* \lesseqqgtr \theta$. Hence, the optimal state independent waiting policy can be found by first finding the optimal policy for the linearized objective function for a given $\theta$ and then finding the optimal $\theta$ using a bisection search. Therefore, the structure of the optimal waiting policy that maximizes $J(\theta)$ is also optimal for the fractional objective. Next, we will present the structure of the optimal waiting policy that maximizes $J(\theta)$.

\begin{theorem}\label{thrm:state_ind_wait}
    Define $p(t)=\sum_{i=1}^S\pi_iP_{i,\hat{X}(t)}(t)$. Then, the optimal waiting function $W^\theta(d)$ that maximizes $J(\theta)$ is,
    \begin{align}
      W^\theta(d)=\argmax_{w\in[0,W_{max}]}\int\limits_{0}^{\infty}(p(t+d)-\theta)\bar{F}^Z(t-w)\dd{t},\label{eqn:wait_sol}
    \end{align}
    where $W_{max}$ is the maximum waiting time allowed and $\bar{F}^z(t)=1-F^Z(t)$.
    \end{theorem}
    \begin{corollary}\label{cor:mono_pt}
      If $p(t)$ is monotonically decreasing in $t$, then, the structure of the optimal waiting policy that maximizes $J(\theta)$ is,
    \begin{align}
      W^{\theta}(d)=\begin{cases}
          W_{max},&\text{if}~~p(\infty)\geq\theta,\\
          0,& \text{if}~~p(d)\leq\theta,\\
          w\in[0,W_{max}],&\text{if}~~p(d)>\theta,p(\infty)<\theta,
      \end{cases}
    \end{align}
    where $w$ is solution to \eqref{eqn:wait_sol}.
\end{corollary}
 
\begin{remark}
    Since the CTMC is ergodic, $p(\infty)$ is well-defined for regular estimators such as MAP or martingale estimators.
\end{remark}

\begin{remark}
    For time-reversible CTMCs,  $P_{i,\hat{X}(t)}(t)=P_{i,i}(t)$  is a continuous monotonically decreasing function  under a martingale estimator. This is true for $p(t)$ as well.
\end{remark}

The proofs of Theorem \ref{thrm:state_ind_wait} and Corollary \ref{cor:mono_pt} is presented in Appendix \ref{appen:thrm:state_ind_wait} and Appendix \ref{appen:cor:mono_pt}, respectively. The third case in Theorem \ref{thrm:state_ind_wait} is trivial and is simply present to convey the possibility that there may exist a non-trivial waiting time in that particular scenario. In Theorem \ref{thrm:thresh_policy}, we show that we can further refine the structure of the waiting policy if the distribution function of $Z$ is absolutely continuous. The proof of Theorem \ref{thrm:thresh_policy} is given in Appendix \ref{appen:thrm:thresh_policy}.

\begin{theorem}\label{thrm:thresh_policy}
    Suppose $p(t)$ is a monotonically decreasing function and the distribution function $F^Z(z)$ is absolutely continuous with respect to the probability density function $f^Z(z)$. Then, the optimal waiting policy  that maximizes $J(\theta)$ has the following threshold structure,
    \begin{align}
        W(d)=\min\{W_{max},(\Gamma-d)^+\},\label{eqn:thresh_1}
    \end{align}
    where $(x)^+=\max\{0,x\}$ and $\Gamma=\sup\{\gamma>0:l(\gamma)\geq\theta\}$ with $l(\gamma)$ defined as,
    \begin{align}
       l(\gamma)=\int\limits_{0}^{\infty}p(t+\gamma)f^Z(t)\dd{t}.\label{eqn:thresh_2}
    \end{align}
    If no such $\Gamma$ exits, then $\Gamma=0$ if $l(0)\leq\theta$ or $\Gamma=\infty$ if $l(d_{max}+W_{max})\geq \theta$ where $d_{max}=\sup\mathcal{D}$.
\end{theorem}

\subsection{Delay-Independent Waiting Policy}\label{sec:delay_ind_wait}
In here, we assert that the waiting function only depends on the most recently received state, and hence, can be reduced to finding  $S$ constants, $W_i$ for $i\in\mathcal{S}$, where $W_i$ represents the waiting time that should be followed, if the most recent reply was from state $i$. One complication that may arise in this setup is that $\bm\phi$ is no longer independent of our waiting policy. Hence, we cannot exploit the same technical intricacies as in Section \ref{sec:state_ind} to find the optimal waiting times analytically. Therefore, we formulate our problem as an SMDP in order to find the optimal $W_i$s. The decision epochs of this SMDP start with the reception of a reply to  a query, where the remote monitor will select a waiting time in the interval $[0,W_{max}]$ based on the state indicated by the reply. This SMDP can be  characterized using the tuple $(\mathcal{S,A,P,R,H)}$ defined below:
\begin{itemize}
    \item The \emph{state space} $\mathcal{S}=\{1,2,\dots,S\}$ is the state of the most recently received reply.
    \item The \emph{action space} $\mathcal{A}=[0,W_{max}]$ is the set of wait times.
    \item The \emph{transition function} $\mathcal{P}:\mathcal{S\cross A\cross S}\to [0,1]$ defines the transition probabilities between states based on the selected action. In particular, $\mathcal{P}(s,a,s')$ denotes the probability of transitioning to state $s'$ if the action $a$ was selected at state $s$. In here,  $\mathcal{P}(i,W_i,j)=\tilde{P}_{i,j}$.
    \item The \emph{reward function} $\mathcal{R}:\mathcal{S\cross A}\to \mathds{R}$ defines the average reward obtained in a  state based on the action selected. In here, $\mathcal{R}(s,a)$ denotes the average reward obtained by selecting action $a$ in state $s$. When in state $i$, if we choose $W_i$ as the waiting time, then $\mathcal{R}(i,W_i)=\e\left[\int\limits_{D'}^{D'+W_i+Z}P_{i,\hat{X}(t)}(t) \dd{t} \right]$.
    \item The \emph{sojourn times} $\mathcal{H}$ defines the average time the process stays in a particular state based on the action selected. In particular, $\mathcal{H}(s,a)$ denotes the average time the process stays in state $s$ if the action $a$ was selected. For example, if we choose the wait time $W_i$ when in state $i$, then $\mathcal{H}(i,W_i)=W_i+\e[Z]$.
\end{itemize}
Since the SMDP is irreducible under any stationary deterministic policy, the optimal waiting policy is a stationary deterministic policy which can be found using the policy iteration algorithm \cite{Ross_CSMDP}.

\subsection{Greedy Waiting Policy}
In here, we consider that the waiting function depends on both the received state $X$ and the observed backward delay. However, in here, we will be maximizing freshness by ignoring the effect of waiting times on the distribution of $\bm \phi$. First, we note that the freshness in \eqref{eqn:fresh_phi} can be bounded as follows,
\begin{align}
    \e[\Delta]\geq\min_i \left\{\frac{\e[g^W(i)]}{\e[Z]+\e[W(i,D)]}\right\}.
\end{align}
Therefore, instead of directly maximizing freshness, we will instead aim to maximize this lower bound. Since the waiting functions for each state is only linked through the distribution $\bm\phi$ which we have ignored in this particular case, we can maximize this lower bound by maximizing each fractional term independently. In particular, for each $i$, we will find the waiting function $W(i,d)$ that maximizes the objective function
$\e[g^W(i)]/(\e[Z]+\e[W(i,D)])$. As before, we will first linearize the objective function and find the structure of the optimal policy that maximizes the linearized objective. Then, this same policy can be translated to the original objective function by finding the optimal linearization parameter through a bisection search. Let the linearized objective function be defined as follows,
\begin{align}
    J_i(\theta)=\e[g^W(i)]-\theta_i(\e[Z]+\e[W(i,D)]),
\end{align}
where $\theta_i$ is the linearization parameter. Then, we can directly translate the results in Theorem \ref{thrm:state_ind_wait} and Theorem \ref{thrm:thresh_policy}  for this problem by replacing the function $p(t)$ with $P_{i,\hat{X}(t)}(t)$.
 
\subsection{Optimal Waiting Policies}
In here, we will modify the SMDP defined in Section \ref{sec:delay_ind_wait} to accommodate waiting times that are dependent on both the received state and the delay of the reply. The state space in this case is two-dimensional with the received state and its subsequent delay being the elements of this state space. Therefore, the curse of dimensionality prohibits the application of this approach to find the optimal waiting times for arbitrary delay distributions. To simplify the state space, we will approximate the delay distribution $D$ with a suitable discrete random variable $\tilde{D}$. Now, the resulting SMDP can be characterized using the tuple $(\mathcal{\tilde{S},\tilde{A},\tilde{P},\tilde{R},\tilde{H})}$ defined below.
\begin{itemize}
    \item The \emph{state space} $\tilde{\mathcal{S}}=(\mathcal{S}\cross\tilde{\mathcal{D}})$, where $\tilde{\mathcal{D}}$ is the support of the discrete random variable $\tilde{D}$.
    \item The \emph{action space} $\tilde{\mathcal{A}}=[0,W_{max}]$ is the set of wait times.
    \item The \emph{transition function} $\tilde{\mathcal{P}}:\mathcal{\tilde{S}\cross \tilde{A}\cross \tilde{S}}\to [0,1]$ with $P((i,d),W(i,d),(j,d'))=\mathds{P}(\tilde{D}=d')\int_YP_{i,j}(d+W(i,d)+y)\,dF^Y(y)$.
    \item The \emph{reward function} $\tilde{\mathcal{R}}:\mathcal{\tilde{S}\cross \tilde{A}}\to \mathds{R}$, where in state $(i,d)$, if we choose $W(i,d)$ as the waiting time then, $\mathcal{R}((i,d),W(i,d))=\e\left[\int\limits_{d}^{d+W(i,d)+Z}P_{i,\hat{X}(t)}(t) \dd{t} \right]$.
    \item The \emph{sojourn times} $\tilde{\mathcal{H}}$ given by $\tilde{\mathcal{H}}((i,d),W(i,d))=W(i,d)+\e[Z]$, if  we choose wait time $W(i,d)$ when in state $(i,d)$.
\end{itemize}
Solving the SMDP using the policy iteration algorithm yields the optimal waiting policy.

\section{Numerical Results}
In this section, we evaluate and compare the performance our waiting policies along with two benchmark policies which we will refer to as ZW policy and CW policy. ZW policy stands for the naive zero-wait policy, where one would immediately send out the next query upon receiving a reply. The CW policy stands for constant wait policy, where we will wait for a fixed amount of time regardless of the observed state or the perceived age of the reply. In here, the fixed wait time will be selected from $[0,W_{max}]$ so that it maximizes the freshness. We will assume that $W_{max}=1.5$ and the martingale estimator is used in all experiments. 

We will refer to our optimal state independent policy, optimal delay independent policy and optimal policy as \emph{state\_ind}, \emph{delay\_ind} and \emph{opt\_wait}. To properly illustrate the difference between the policies and to simplify the SMDP formulations, we present three examples under discrete delay distributions. Additionally, in all experiments, we will be considering a binary CTMC whose transition rate from state $1$ to $2$ is $\alpha$ and from state $2$ to state $1$ is $\beta$. Binary CTMCs will help to further reduce the state space in addition to being time reversible. We will explain the dynamics of the waiting policies in detail for the first experiment and will only highlight the key differences in the other two experiments.

\begin{figure}
    \centering
    \includegraphics[width=\custmsz\columnwidth]{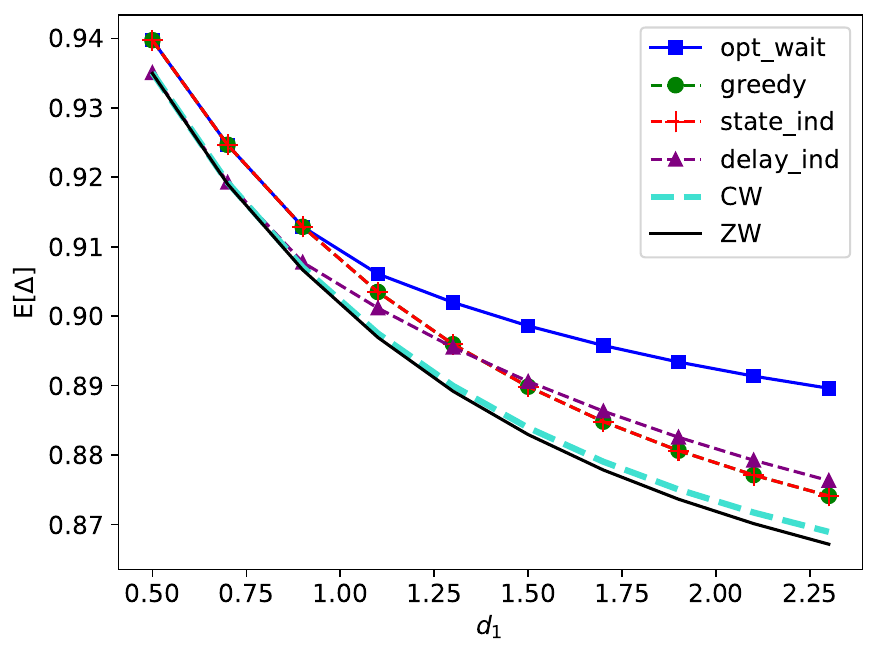}
    \caption{Variation of mean binary freshness for a binary CTMC  with $\alpha=1$, $\beta=0.1$, $Y=0$ and $D=\{0,d_1\}$ with probabilities $\{0.5,0.5\}$.}
    \label{fig:var_d_1}
\end{figure}

In the first experiment, we will be considering that $\alpha=1$ and $\beta=0.1$, thus making state $1$ a less probable or less stable state, and state $2$ a high probable or a more stable state.  For this experiment, to clearly highlight the differences in the waiting policies, we will assume that there is no forward delay ($Y=0$) and $D$ is a discrete random variable which takes values $0$ and $d_1$ with probability $0.5$. We will observe how the freshness metric will evolve as we vary the parameter $d_1$ of the backward delay. As illustrated in Fig.~\ref{fig:var_d_1}, there is a noticeable gain when appropriately waiting before sending out the next query. Moreover, we see that for lower $d_1$ values, the \emph{state\_ind} policy closely follows the optimal curve and is performing better than the \emph{delay\_ind} policy. However, as the delay is increased \emph{delay\_ind} policy exhibits better freshness values. This is because when there are significant delays in the backward channel, the state of the update is more important than the perceived delay of the update since one state is less stable than the other. Additionally, we see that both the \emph{state\_ind} and \emph{delay\_ind} policies have significant gains over the ZW and CW policies for higher delay values.

In here, we notice that the \emph{state\_ind} policy would wait for some time if the perceived age of the estimator is low and would immediately sample if it is high. This is to avoid immediately sampling the less probable state upon receiving a zero delay sample. In the \emph{delay\_ind} policy, we observe that, at higher $d_1$ values, this policy would wait the maximum waiting period when in the high probable state whereas it would only enforce a short waiting period in the less probable state. Thus, for higher delay values, the system prefers to latch on to the high probable state as its estimate for as long as possible, and would like to quickly update if it receives a sample from the less probable state. At lower $d_1$ values, this policy is similar to the CW policy. In the \emph{opt\_wait} policy, the system would have a non-zero waiting time if it receives a sample from the less probable state with zero delay, and would wait the maximum waiting time if it receives a sample from the high probable state with a non-zero delay. For every other combination, it would have zero wait time. This essentially combines the two contrasting strategies of the \emph{state\_ind} and \emph{delay\_ind} policies. 

\begin{figure}
    \centering
    \includegraphics[width=\custmsz\columnwidth]{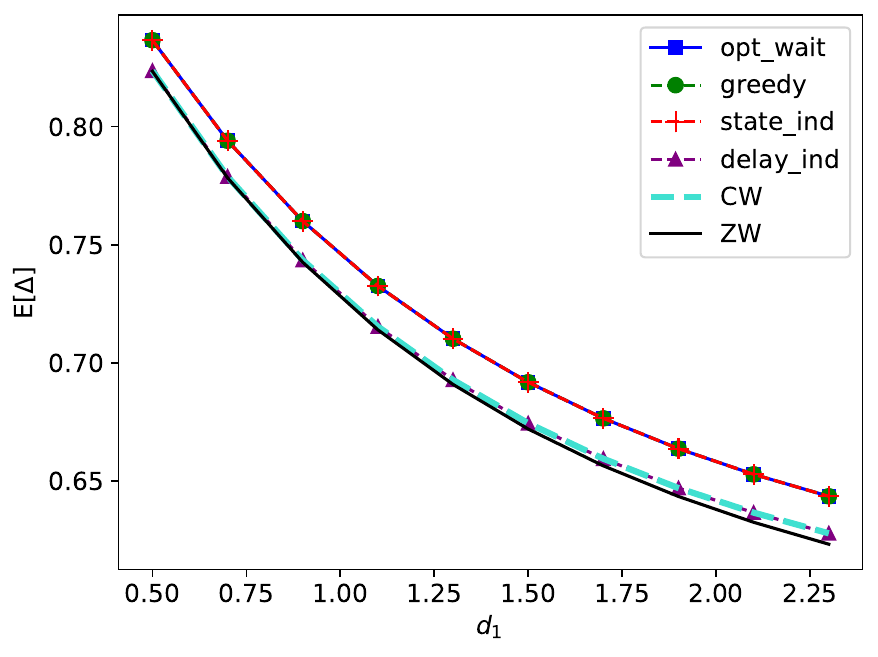}
    \caption{Variation of mean binary freshness for a binary CTMC  with $\alpha=0.6$, $\beta=0.4$, $Y=0$ and $D=\{0,d_1\}$ with probabilities $\{0.5,0.5\}$.}
    \label{fig:var_d_2}
\end{figure}

\begin{figure}
    \centering
    \includegraphics[width=\custmsz\columnwidth]{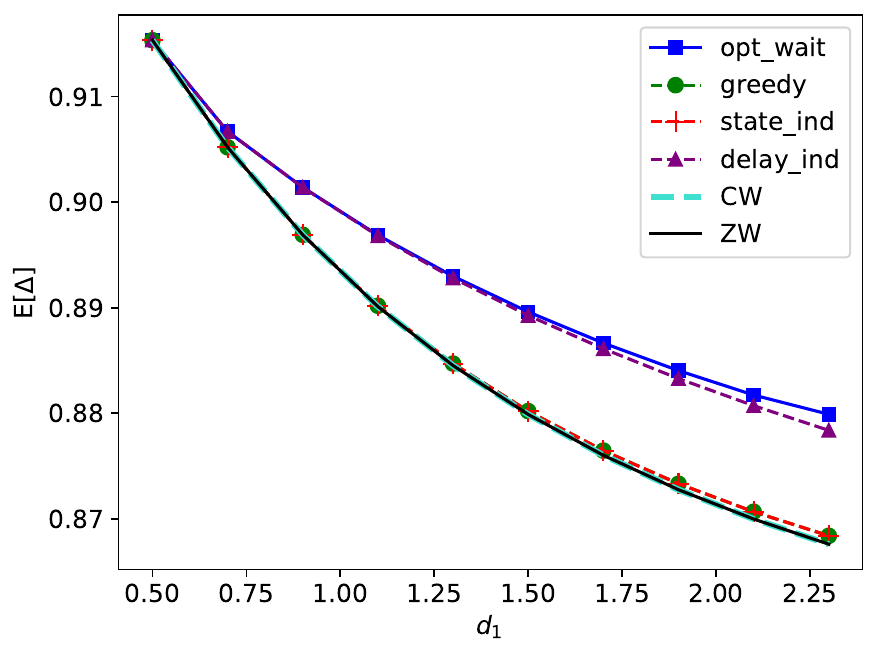}
    \caption{Variation of mean binary freshness for a binary CTMC  with $\alpha=1$, $\beta=0.1$, $Y= \{0.3,0.5,1\}$ with probabilities $\{0.3,0.3,0.4\}$ and $D=\{0,d_1\}$ with probabilities $\{0.5,0.5\}$.}
    \label{fig:var_d_3}
\end{figure}

In the second experiment, we consider the same setup as in the first experiment but with $\alpha=0.6$ and $\beta=0.4$. Here, both states are stable. As seen in Fig.~\ref{fig:var_d_2}, when both the states are stable, the state of the received update is less important, which is the reason why the \emph{state\_ind} policy is on par with the \emph{opt\_wait} policy and outperforms the \emph{delay\_ind} policy. 

In the third experiment, to demonstrate the effect of non-zero query delays, we consider the same step up as in the first experiment and consider that $Y$ takes values $\{0.3,0.5,1\}$ with probabilities $\{0.3,0.3,0.4\}$. As depicted in Fig.~\ref{fig:var_d_3}, when the query delay is also non-zero, we see that the \emph{delay\_ind} policy is close to the optimal policy and outperforms the \emph{state\_ind} policy across all $d_1$ values. 

Finally, we observe that across all three experiments, the greedy policy and the \emph{state\_ind} policy are closely aligned. The actual technical reason  behind this observation remains to be uncovered.

\begin{remark}
    While our waiting policies enhance freshness, we have noticed in the presence of delays, the system tries to avoid sampling less stable states so as to maximize freshness. This creates an observation bias towards more stable states  in our estimates. While adding different semantic penalties for each observed state may be a simple solution to mitigate this bias, we note that by doing so, one would never truly track the actual state of the system.
\end{remark}

\section{Conclusion}
In this work, we have studied the query-based sampling procedure under generic delay distributions and have shown that by appropriately waiting for some time before sending out queries one can enhance the MBF. We have shown that under certain assumptions a threshold structure emerges and we have given an SMDP formulation to find the optimal waiting policy. Future directions of research in this avenue involve expanding this analysis for metrics such as AoII.

\appendices
\section{Proof of Theorem \ref{thrm:fresh_gen}}\label{appen:fresh_gen}
Let $Y_k$ be the forward delay for the $k$th query and let $D_k$ be the backward delay for the same query. Let $W(i,d)$ denote the bounded waiting time function which represents the waiting time before sending the next query upon receiving an outdated reply with age $d$ and state $i$. Let $T_k$ be the time at which the monitor received the reply for the $k$th query with $T_0=1$. Then, we have that $T_{k}-T_{k-1}=W_k+Y_k+D_k$, where $W_k=W(X_{k},D_{k-1})$ with $X_{k}=X(T_{k-1}-D_{k-1)}$ being the status of the most recent update. Now, the MBF can be represented using these forward and backward delays as follows,
\begin{align}
    \e[\Delta]&=\lim_{T\to\infty}\frac{1}{T}\e\left[\int_0^T\mathds{1}\{X(t)=\hat{X}(t)\}\dd{t}\right]\\
    &=\lim_{n\to\infty}\frac{1}{T_n}\e\left[\int_0^{T_n}\mathds{1}\{X(t)=\hat{X}(t)\}\dd{t}\right]\label{eqn:fresh_gen_1}\\
    &=\lim_{n\to\infty}\frac{\frac{1}{n}\sum_{k=1}^n\e\left[\int_{T_{k-1}}^{T_k}\mathds{1}\{X(t)=\hat{X}(t)\}\dd{t}\Big |X_k\right]}{\frac{1}{n}\sum_{k=1}^nW_k+Y_k+D_k}\label{eqn:fresh_gen_2}\\
   &= \frac{\lim_{n\to\infty}\frac{1}{n}\sum_{k=1}^n\e\left[\int_{T_{k-1}}^{T_k}\mathds{1}\{X(t)=\hat{X}(t)\}\dd{t}\Big |X_k\right]}{\lim_{n\to\infty}\frac{1}{n}\sum_{k=1}^nW_k+Y_k+D_k}.
\end{align}
For brevity, we have removed the $\limsup$ notation since the subsequent limits exit. In the above equations, \eqref{eqn:fresh_gen_1} holds almost surely and in \eqref{eqn:fresh_gen_2} the expectation is with respect to the randomness of the process $X(t)$ between update times. Next, note that,
\begin{align}
    &\e\left[\int\limits_{T_{k-1}}^{T_k}\mathds{1}\{X(t)=
    \hat{X}(t)\}\dd{t}\Big |X_k\right]\nonumber\\
    &=\e\left[\int\limits_{D_{k-1}}^{D_{k-1}+W_k+Y_k+D_k}\hspace{-2em}\mathds{1}\{X(t)=\hat{X}(t)\}\dd{t}\Big |X(0)=X_k\right].
\end{align}
 Let $h_k=h(X_{k},D_{k-1},Y_k,D_k)=W_k+Y_k+D_k$  and $g_k=g(X_{k},D_{k-1},Y_k,D_k)=\e\left[\int_{T_{k-1}}^{T_k}\mathds{1}\{X(t)=\hat{X}(t)\}\,dt\Big |X_k\right]$. We have $g_k\leq h_k$. Since $W$ is a bounded waiting function, we have that both these functions of random variables have finite expectation. Moreover, note that $X_{k},D_{k-1},Y_k,D_k$ are all independent from each other. Further, we have that, $X_{k-1}$ converge to some random variable $X$ over the support $\mathcal{S}$ since the sampling process is stationary. Additionally, for discrete random variables with finite support convergence in distribution implies convergence in the mean. Therefore, from Birkhoff's ergodic theorem \cite{koralov_sinai}, we have that,
\begin{align}
    \lim_{n\to \infty}\frac{1}{n}\sum_{k=1}^ng_k&\to\e[g(X,D',D,Y)],\\
    \lim_{n\to \infty}\frac{1}{n}\sum_{k=1}^nh_k &\to \e[h(X,D',D,Y)],
\end{align}
where $D'$ is an independent copy of the random variable $D$ and represents the perceived age of the reply. Next, notice that, by conditioning on the realizations of $X=x$, $D'=d'$, $Y=y$ and $D=d$, and then taking expectations with respect to randomness of the process $X(t)$, we have,
\begin{align}
    &\e\left[\int\limits_{d'}^{d'+W(x,d')+y+d}\hspace{-2em}\mathds{1}\{X(t)=\hat{X}(t)\}\dd{t}\Big |X(0)=x\right]\\
    &=\int\limits_{d'}^{d'+W(x,d')+y+d}\hspace{-2em}\e\left[\mathds{1}\{X(t)=\hat{X}(t)\}\Big |X(0)=x\right]\dd{t}\\
    &=\int\limits_{d'}^{d'+W(x,d')+y+d}\hspace{-2em}P_{x,\hat{X}(t)}(t)\dd{t}.
\end{align}
Now, accounting for the randomness of $X$, $D'$, $Y$ and $D$ yields,
\begin{align}
    \e[g(X,D',D,Y)]=\e\left[\int\limits_{D'}^{D'+W(X,D')+Z}\hspace{-2em}P_{X,\hat{X}(t)}(t)\dd{t}\right].
\end{align}
To find the distribution of the random variable $X$, we note that that the sequence of random variables $X_k$ form a discrete time Markov chain (DTMC). Let $\tilde{P}_{i,j}$ denote the probability $\mathds{P}[X_{k}=j|X_{k-1}=i]$. Then, $\tilde{P}_{i,j}$ will be given by,
\begin{align}
    \tilde{P}_{i,j}=\int_\mathcal{D}\int_\mathcal{Y}P_{i,j}(d+W(i,d)+y)\dd{F^Y(y)}\dd{F^D(d)}.
\end{align}
Since the DTMC is irreducible and aperiodic, its stationary distribution exists. Let $\bm\phi=\{\phi_1,\phi_2,\dots,\phi_S\}$ denote this stationary distribution. Therefore, we have that the distribution of $X$ is $\bm\phi$ and it satisfies $\bm\phi \tilde{P}=\bm\phi$, where $\tilde{P}$ is a stochastic matrix whose entries are $\tilde{P}_{i,j}$. 

\section{Proof of Corollary \ref{cor:fresh_gen}}\label{appen:cor_fresh_gen}
For fixed realizations of $D'=d$ and $X=i$, we have, 
\begin{align}
    &\e[g(i,d,D,Y)]\nonumber\\
    &=\e\left[\int\limits_{d}^{d+W(i,d)+Z}P_{i,\hat{X}(t)}(t)\dd{t}\right]\\
    &=\e\left[\int\limits_{0}^{W(i,d)+Z}P_{i,\hat{X}(t+d)}(t+d)\dd{t}\right]\\
    &=\e\left[\int_{0}^{\infty}P_{i,\hat{X}(t+d)}(t+d)\dd{t}\mathds{1}\{Z+W(i,d)>t\}\dd{t}\right]\label{eqn:exp_Z}\\
    &=\int_{0}^{\infty}P_{i\hat{X}(t+d)}(t+d)(1-F^{Z}(t-W(i,d))\,\dd{t}.\label{eqn:exp_z_interchange}
\end{align}
where \eqref{eqn:exp_z_interchange} was obtained by taking the expectation inside the integral by the virtue of Tonelli's theorem \cite{zygmund}. Now, accounting for the randomness of $X$ and $D'$, we get,
\begin{align}
    \e[g(X,D',D,Y]=\sum_{i=1}^S\phi_i\int_{\mathcal{D}}\e[g(i,d,D,Y)]\dd{F^D(d)}.
\end{align}
Similarly, we have,
\begin{align}
    \e[h(X,D',D,Y]=\e[Z]+\sum_{i=1}^S\phi_i\e[W(i,D)].
\end{align}
\section{Proof of Lemma \ref{lem:state_ind}}\label{appen:lem:state_ind}
Note that the distribution $\bm\pi$ satisfies the following,
\begin{align}
    &\sum_{i=1}^S\pi_i\tilde{P}_{ij}\nonumber
    \\&=\sum_{i=1}^S\pi_i\int_D\int_YP_{ij}
    (d+W(d)+y)\dd{F^Y(y)}\dd{F^D(d)}\\
    &=\int_D\int_Y\underbrace{\sum_{i=1}^S\pi_iP_{ij}
    (d+W(d)+y)}_{\pi_j}\dd{F^Y(y)}\dd{F^D(d)}\\
    &=\pi_j.
\end{align}
Moreover, $\sum_{i=1}^S\pi_i=1$. Since $\bm\phi$ is unique, we have  $\bm\phi=\bm\pi$.

\section{Proof of Theorem \ref{thrm:state_ind_wait}}\label{appen:thrm:state_ind_wait}
 We have,
\begin{align}
    J(\theta)&=\e\left[\int\limits_{D'}^{D'+W(D')+Z}\hspace{-2em}P_{X,\hat{X}(t)}(t)\dd{t}-\theta (W(D')+Z)\right]\\
    &=\e\left[\int\limits_{D'}^{D'+W(D')+Z}\hspace{-2em}P_{X,\hat{X}(t)}(t)\dd{t}-\int\limits_{0}^{W(D')+Z}\theta \dd{t}\right]\\
    &=\e\left[\int\limits_{0}^{W(D')+Z}\hspace{-1em}P_{X,\hat{X}(t+D')}(t+D')\dd{t}-\int\limits_{0}^{W(D')+Z}\theta \dd{t}\right]\\
    &=\e\left[\int\limits_{0}^{W(D')+Z}\hspace{-1em}P_{X,\hat{X}(t+D')}(t+D')-\theta \dd{t}\right]\\
    &=\e\left[\e\left[\int\limits_{0}^{W(D')+Z}\hspace{-1em}P_{X,\hat{X}(t+D')}(t+D')-\theta \dd{t}\Bigg|D'\right]\right].
\end{align}
Therefore, to maximize $J(\theta)$, it is sufficient for $W(D')$ to be chosen so as to maximize $\e\left[\int\limits_{0}^{W(D')+Z}\hspace{-1em}P_{X,\hat{X}(t+D')}(t+D')-\theta \dd{t}\Bigg|D'\right]$. Now, suppose $D'=d$. Therefore, we have,
\begin{align}
    &\e\left[\int\limits_{0}^{W(d)+Z}\hspace{-1em}P_{X,\hat{X}(t+d)}(t+d)-\theta \dd{t}\right]\nonumber\\
    &=\sum_{i=1}^S\pi_i\e\left[\int\limits_{0}^{W(d)+Z}\hspace{-1em}P_{i,\hat{X}(t+d)}(t+d)-\theta \dd{t}\right]\\
    &=\e\left[\int\limits_{0}^{W(d)+Z}\sum_{i=1}^S\pi_iP_{i,\hat{X}(t+d)}(t+d)-\theta \dd{t}\right]\\
&=\e\left[\int\limits_{0}^{W(d)+Z}\hspace{-1em}p(t+d)-\theta \dd{t}\right].
\end{align}
Now, similar to  \eqref{eqn:exp_Z}, we can further refine the above expression as,
\begin{align}
  &\e\left[\int\limits_{0}^{W(d)+Z}\hspace{-1em}p(t+d)-\theta \dd{t}\right]\nonumber\\
  &=\int\limits_{0}^{\infty}(p(t+d)-\theta)\left(1-F^Z(t-W(d))\right)\dd{t}\label{eqn:max_pt}.
\end{align}
Therefore, the optimal waiting time is the argument that maximizes \eqref{eqn:max_pt}.

\section{Proof of Corollary \ref{cor:mono_pt}}\label{appen:cor:mono_pt}
We have that $p(t)$ is monotonically decreasing.  Therefore, if $p(\infty)\geq\theta$, we have that $p(t+d)-\theta\geq0$ for all $t>0$. Now, note that $1-F^Z(t-W(d))=1$ for $t\leq W(d)$ and monotonically decreases to zero for $t>W(d)$. Therefore, to maximize \eqref{eqn:max_pt}, we want to maximize the region that $1-F^Z(t-W(d))=1$. Thus, we choose $W(d)=W_{max}$ in this case. On the other hand, if $p(d)\leq\theta$, then we have that $p(t+d)-\theta<0$. Therefore, in order to maximize \eqref{eqn:max_pt}, we need  $1-F^Z(t-W(d))$ to approach zero faster. Thus, $W(d)=0$ is the optimal choice for this particular case. If $p(d)>\theta$ and $p(\infty)<\theta$, then $p(t+d)-\theta$ is initially positive and then eventually becomes negative. Therefore, we need to maximize the value of $1-F^Z(t-W(d))$ in the positive region and allow it to decay in the negative region. Hence, there may exist a non trivial $W(d)$ that maximizes \eqref{eqn:max_pt} for this particular case.

\section{Proof of Theorem \ref{thrm:thresh_policy}}\label{appen:thrm:thresh_policy}
Let $L(w)=\int\limits_{0}^{\infty}(p(t+d)-\theta)\left(1-F^Z(t-w)\right)\dd{t}$ and $U(w)=\int\limits_{0}^{\infty}(p(t+d)-\theta)f^Z(t-w)\dd{t}$. Since $p(t+d)-\theta$ is a bounded function and $f^Z$ is a probability density, we have from Fubini's theorem \cite{zygmund},
\begin{align}
    \int_{0}^wU(u)\dd{u}&=\int\limits_{0}^{\infty}(p(t+d)-\theta)\int_{0}^{w}f^Z(t-u)\dd{u}\dd{t}\\
    &=\int\limits_{0}^{\infty}(p(t+d)-\theta)(1-F^Z(t-w))\dd{t}\nonumber\\
    &\quad-\int\limits_{0}^{\infty}(p(t+d)-\theta)(1-F^Z(t))\dd{t}\\
    &=L(w)-L(0).
\end{align}
Moreover, from the continuity of translation theorem, we have that $U(w)$ is continuous with respect to $w$. Therefore, from the fundamental theorem of calculus we have,
\begin{align}
    \dv{L(w)}{w}&=\int\limits_{0}^{\infty}(p(t+d)-\theta)\dv{\left(1-F^Z(t-w)\right)}{w}\dd{t}\\
    &=\int\limits_{0}^{\infty}(p(t+d)-\theta)f^Z(t-w)\dd{t}\\
    &=\int\limits_{0}^{\infty}(p(t+w+d)-\theta)f^Z(t)\dd{t}.\label{eqn:thresh_fun}
\end{align}
Let $w^*$ be the optimal wait time. Since $p(t)$ is monotonically decreasing, we have that $\dv{L(w)}{w}$ is monotonically decreasing. Therefore, if $\dv{L(w)}{w}\big|_{w=0}\leq 0$, then the optimal waiting time $w^*=0$. If $\dv{L(w)}{w}\big|_{w=W_{max}}\geq0$, then $w^*=W_{max}$. If $\dv{L(w)}{w}\big|_{w=W_{max}}<0<\dv{L(w)}{w}\big|_{w=0}$, then $w^*$ is achieved when $\dv{L(w)}{w}=0$. This results in the following condition,
\begin{align}
    \int\limits_{0}^{\infty}p(t+w^*+d)f^Z(t)\dd{t}=\theta.
\end{align}
Let $l(\gamma)=\int\limits_{0}^{\infty}p(t+\gamma)f^Z(t)\dd{t}$. If $l(0)\leq \theta$, then for any delay $d$, we would have $\dv{L(w)}{w}\leq0$, since $p(t)$ is monotonically decreasing. Therefore, in this case the optimal waiting policy is the zero wait policy (i.e., $W(d)=0~\forall d$). If $l(W_{max}+d_{max})\geq0$, since $p(t)$ is monotonically decreasing, we have $\dv{L(w)}{w}\geq0~\forall d$. Hence the optimal waiting policy is $W(d)=W_{max}~\forall d$. Finally, since $p(t+\gamma)f^Z(t)$ is an integrable function, as before, from the continuity of translation theorem, we have that $l(\gamma)$ is continuous. As a result, if $l(0)>0$  and $l(W_{max}+d_{max})<0$, then $\exists \gamma>0$ such that $l(\gamma)=\theta$. Let $\Gamma=\sup{\gamma>0:l(\gamma)=\theta}$. Therefore, in this case, if our delay $d<\Gamma$, then $w^*=\Gamma-d$. Otherwise the optimal waiting time is zero. These three scenarios have been concisely summarized in \eqref{eqn:thresh_1} and \eqref{eqn:thresh_2}. 


\bibliographystyle{unsrt}
\bibliography{refs}

\end{document}